\documentclass[aps,prl,twocolumn,showpacs,superscriptaddress,10pt]{revtex4-1}

\usepackage{graphicx}
\usepackage{color}
\usepackage{amsmath}
\usepackage{amsfonts}
\usepackage{bm}

\newcommand{\bk}{\mathbf{k}}
\newcommand{\abs}[1]{\ensuremath{\lvert #1 \rvert}}
\newcommand{\vect}[1]{\mathbf{#1}}
\newcommand{\ket}[1]{\ensuremath{|#1\rangle}}

\begin{document}

\title{Majorana Modes in Driven-Dissipative Atomic Superfluids With Zero Chern Number}

\author{C.-E. Bardyn}
\affiliation{Institute for Quantum Electronics, ETH Zurich, 8093 Zurich, Switzerland}

\author{M. A. Baranov}
\affiliation{Institute for Quantum Optics and Quantum Information of the Austrian Academy of Sciences, A-6020 Innsbruck, Austria}
\affiliation{Institute for Theoretical Physics, University of Innsbruck, A-6020 Innsbruck, Austria}
\affiliation{NRC ``Kurchatov Institute'', Kurchatov Square 1, 123182 Moscow, Russia}

\author{E. Rico}
\affiliation{Institute for Quantum Optics and Quantum Information of the Austrian Academy of Sciences, A-6020 Innsbruck, Austria}
\affiliation{Institute for Theoretical Physics, University of Innsbruck, A-6020 Innsbruck, Austria}

\author{A. \.Imamo\u glu}
\affiliation{Institute for Quantum Electronics, ETH Zurich, 8093 Zurich, Switzerland}

\author{P. Zoller}
\affiliation{Institute for Quantum Optics and Quantum Information of the Austrian Academy of Sciences, A-6020 Innsbruck, Austria}
\affiliation{Institute for Theoretical Physics, University of Innsbruck, A-6020 Innsbruck, Austria}

\author{S. Diehl}
\affiliation{Institute for Quantum Optics and Quantum Information of the Austrian Academy of Sciences, A-6020 Innsbruck, Austria}
\affiliation{Institute for Theoretical Physics, University of Innsbruck, A-6020 Innsbruck, Austria}

\begin{abstract}

We investigate dissipation-induced $p$-wave paired states of fermions in two dimensions and show the existence of spatially separated Majorana zero modes in a phase with vanishing Chern number. We construct an explicit and natural model of a dissipative vortex that traps a single of these modes, and establish its topological origin by mapping the problem to a chiral one-dimensional wire where we observe a non-equilibrium topological phase transition characterized by an abrupt change of a topological invariant (winding number). We show that the existence of a single Majorana zero mode in the vortex core is intimately tied to the dissipative nature of our model. Engineered dissipation opens up possibilities for experimentally realizing such states with no Hamiltonian counterpart.

\end{abstract}

\maketitle


The search for topological phases of matter in which elementary excitations exhibit non-Abelian statistics have brought two-dimensional (2D) $p$-wave paired superfluids and superconductors to the forefront of theoretical and experimental condensed-matter research~\cite{Read00, Ivanov01, Kitaev03, Kitaev06, Nayak08}. The bulk of these systems is fundamentally intriguing in that it reveals physics beyond the Landau paradigm: different phases are characterized by distinct values of a nonlocal, \emph{topological} order parameter known as the Chern number, and phase transitions occur whenever the topology changes, signaled by discontinuities in this integer-valued topological invariant. In topologically non-trivial phases corresponding to an \emph{odd} Chern number, vortices with odd vorticity have been predicted to carry \emph{unpaired} Majorana fermions, and to exhibit non-Abelian exchange statistics as a result~\cite{InterferometryProposals}.

In this Letter, we explore the concept of topological order and its connection to the edge physics in a non-equilibrium scenario based on engineered dissipation~\cite{Diehl08, Verstraete09}. Prior work on quantum-state engineering in driven-dissipative systems has shown that topologically non-trivial states of many-body Hamiltonians can also be prepared as steady states of a dissipative dynamics~\cite{Diehl11}. In contrast, we demonstrate that dissipation can lead to a novel manifestation of topological order with no Hamiltonian counterpart. Specifically, we show that spatially separated Majorana zero modes (MZMs) can be obtained in a 2D, dissipation-induced $p$-wave paired phase of spin-polarized fermions with \emph{vanishing} Chern number. Remarkably, a phase whose topological nature is seemingly trivial --according to the standard diagnostic tool provided by the Chern number-- can therefore exhibit phenomenological features characteristic of a non-trivial one, which ultimately leads to the counterintuitive fact that vortices with odd vorticity may obey non-Abelian exchange statistics in a bulk with zero Chern number.

We demonstrate these results in a simple model motivated by an implementation scheme based on cold atoms and optical vortex imprinting~\cite{Greiner11}, where fermion parity is microscopically conserved. We show that they hold over an extended parameter range, in which we identify a topologically non-trivial phase missed by Chern number considerations. Critical points are revealed upon introducing a vortex, in which case we establish the phenomenology of a non-equilibrium topological phase transition characterized by (i) a discontinuity in a topological invariant (winding number), (ii) divergent length and time scales~\cite{Verstraete09, Diehl10, Eisert10, Hoening11}, and (iii) a divergent localization length associated with a MZM bound to the vortex core.

The basic mechanism behind our findings relies on the fact that the introduction of a (dissipative) vortex changes the system in two crucial respects: it modifies its topology, as argued in Refs.~\cite{Tewari07, Hou07}, and imposes specific (dissipative) boundary conditions. Here we show that our model of a vortex with odd vorticity can be mapped to a 1D chiral fermion problem~\cite{Kitaev00} characterized by a non-trivial topological invariant (winding number) $\nu_{\text{1D}} = 2$ despite a vanishing bulk Chern number. In such a situation, bulk-edge correspondence arguments~\cite{Hatsugai93, Kitaev06, Ryu10, Gurarie11} suggest the existence of a \emph{pair} of MZMs in the vortex core. However, owing to the dissipative boundary conditions imposed by the geometry of the vortex core alone --which underpins the universal nature of our findings-- a \emph{single} MZM only is found in the core. This phenomenon crucially relies on dissipation, and therefore has no Hamiltonian counterpart. It shows that the potentially harmful effect of dissipation on MZMs~\cite{Chamon11, Trauzettel11, Loss12} need not be entirely destructive, but may instead give rise to intriguing novel effects.

\emph{Dissipative framework} -- We consider a system of $N$ fermionic sites $a_i^\dagger, a_i$ evolving under a purely dissipative dynamics governed by a Lindblad master equation
\begin{align} \label{eqn:masterEquation}
	\partial_t \rho = \kappa \sum_{i = 1}^N \left( L_i \rho L_i^\dagger - \tfrac{1}{2} \{ L^\dagger_i  L_i, \rho \} \right),
\end{align}
where $\rho$ is the system density matrix, $\kappa$ the damping rate, and $L_i$ are Lindblad operators which are linear in the fermionic operators. The steady state of such dynamics is pure if and only if the Lindblad operators form a set of anticommuting operators, i.e. $\{ L_i, L_j \} = 0$ for all $i, j = 1, \ldots, N$. If so, it can be identified with the ground state of the \emph{parent Hamiltonian} $H_{\text{parent}} \equiv \sum_i L^\dagger_i L_i$. Although purity is not required for the existence of topological order~\cite{Diehl11}, we will only encounter steady states that are pure in the bulk, whose (bulk) topological properties can be inferred from $H_{\text{parent}}$ alone. In our dissipative setting, the analog of a gap in a Hamiltonian spectrum is a \emph{dissipative} gap in the Liouvillian spectrum, which \emph{dynamically} isolates the subspaces corresponding to the bulk and edge modes, thereby providing the counterpart of a gap protection through the quantum Zeno effect~\cite{Beige00}. Most importantly, the counterpart of topological ground-state degeneracy is the existence of a nonlocal decoherence-free subspace associated with zero-damping Majorana modes $\gamma = \gamma^\dagger$ which satisfy the orthogonality condition $\{ L_i, \gamma \} = \{ L^\dagger_i, \gamma \} = 0$ for all $i$ (see Supplementary Information (SI)). Clearly, this condition is more restrictive than the one for a zero-energy Majorana mode of a Hamiltonian, which reads (for $H_{\text{parent}}$) $[ H_\text{parent}, \gamma ] = \sum_i ( L_i^\dagger \{ L_i, \gamma \} - \{ L^\dagger_i, \gamma \} L_i ) = 0$. The crucial difference stems from the first ``recycling'' term on the right-hand side of Eq.~\eqref{eqn:masterEquation}, and is the reason why dissipation can crucially modify the Majorana physics found in the Hamiltonian context.

\emph{The model} -- We consider a square-lattice system driven by so-called ``cross'' Lindblad operators defined as the following quasi-local linear superposition of fermionic creation and annihilation operators $a_i^\dagger, a_i$:
\begin{align} \label{eqn:crossLindbladOperators}
	\begin{split}
		L_i \equiv C_i^\dagger + \, \alpha \, e^{\mathrm{i} \phi} A_i = \beta \, a^\dagger_i \; + \; & ( a^\dagger_{i_1} + \hphantom{\mathrm{i}} a^\dagger_{i_2} + \hphantom{\mathrm{i}} a^\dagger_{i_3} + \hphantom{\mathrm{i}} a^\dagger_{i_4} ) \\
		\; + \; \alpha \, e^{\mathrm{i} \phi} & ( a_{i_1} + \mathrm{i} a_{i_2} - \hphantom{\mathrm{i}} a_{i_3} - \mathrm{i} a_{i_4} ),
	\end{split}
\end{align}
where $\beta \in \mathbb{R}$, $\alpha > 0$, $\phi \in [ 0, 2 \pi )$, and $i_1$, $i_2$, $i_3$ and $i_4$ are the four clockwise-ordered nearest-neighboring sites of $i$. The creation and annihilation parts of $L_i$ have $s$- and $p$-wave symmetries, respectively. The dissipative dynamics generated by such Lindblad operators can be obtained, for example, as the long-time limit of a microscopically number-conserving (quartic) dissipative dynamics generating phase-locked paired states (see SI). In that context, the global relative phase $\phi$ between the creation and annihilation parts of $L_i$ emerges through spontaneous breaking of the global $U(1)$ symmetry, and the relative strength $\alpha$ is determined by the average particle number. The dimensionless parameter $\beta$, on the other hand, can be used to tune the system across phase transitions, as will be shown below.

The steady-state bulk properties of the system are most easily revealed in the infinite-size limit. Defining coefficients $u_{ij}, v_{ij}$ such that $L_i = \sum_j (u_{ij} a_j + v_{ij} a^\dagger_j)$, the momentum-space cross Lindblad operators take the form $L_\bk = u_\bk a_\bk + v_\bk a^\dagger_{-\bk}$, where $u_\bk, v_\bk$ are the Fourier transforms of $u_{ij}, v_{ij}$. Properly normalized, they become Bogoliubov quasiparticle operators associated with damping rates $\kappa_\bk = \kappa \, \mathcal{N}_\bk$, where $\mathcal{N}_\bk = \xi^\dagger_\bk \xi_\bk$ with $\xi_\bk^T = (u_\bk, v_\bk)$. One can easily verify that the cross Lindblad operators form a complete set of anticommuting operators, so that the system is driven into a unique and pure complex $p$-wave paired state $\ket{\Omega}$ defined by the condition $L_\bk \ket{\Omega} = 0$ (for all $\bk$) and fully characterized by the real vector $\vect{n}_\bk = \xi^\dagger_\bk \boldsymbol{\sigma} \xi_\bk$, where $\boldsymbol{\sigma}$ is a vector of Pauli matrices. Since (i) this steady state is pure (i.e. $\abs{\vect{n}_\bk} = 1$ for all $\bk$) and (ii) time-reversal symmetry is broken due to the complex $p$-wave nature of the state~\cite{Footnote1}, the topological invariant relevant to that case coincides with the Chern number $\nu_{\text{2D}}$ commonly used in 2D Hamiltonian systems (see e.g. Ref.~\cite{Volovik88}). Here we find that the Chern number vanishes~\cite{Footnote2}, namely,
\begin{align} \label{chernNumber}
	\nu_{\text{2D}} \equiv \tfrac{1}{4\pi} \int_{\mathrm{BZ}} d^2 \vect{k} \; \vect{n}_\bk \cdot ( \partial_{k_x} \vect{n}_\bk \times \partial_{k_y} \vect{n}_\bk ) = 0
\end{align}
(where BZ stands for ``Brillouin zone'') for all values of $\beta$ except at the isolated points $\beta = 0$ and $\beta = \pm 4$ where the dissipative gap closes (see SI). Since there is no extended parameter range with non-trivial topological order, one naively expects topological features such as isolated MZMs to be absent when edges or vortices are introduced. We will show below that such conclusions are premature: single, unpaired MZMs are generally found in the parameter range $0 < \abs{\beta} < 4$ when dissipative vortices with odd vorticity are introduced.

Physically, the special values $\beta = 0$, $\pm 4$ appear as critical points since the dissipative gap closes at these values, leading to divergent length and time scales characteristic of a second-order phase transition~\cite{Eisert10, Hoening11}. However, the symmetry of the steady state (encoded in $\vect{n}_\bk$) and the value of the topological invariant $\nu_{\text{2D}}$ both are identical in the neighborhood of those points. It thus seems as if the apparent critical behavior can neither be traced to a conventional phase transition (with broken symmetry), nor to a topological one (with a discontinuity in the topological invariant). Below we will show that the introduction of a vortex makes it possible to define another topological invariant. This will allow us to identify $\beta = \pm 4$ as genuine critical points associated with topological phase transitions.

\emph{Introducing a dissipative vortex} -- We now introduce a \emph{dissipative} vortex, by modifying the annihilation part $A_i = \sum_j u_{ij} a_j$ of the above cross Lindblad operators. More specifically, we replace the translation-invariant coefficients $u_{ij} \equiv \tilde{u}_{ij}$ considered so far by position-dependent coefficients of the form $u_{ij} = f(r_j) e^{- \mathrm{i} \ell \varphi_j} \tilde{u}_{ij}$, where $(r_j, \varphi_j)$ are polar coordinates defined with respect to a particular site $i_0$ chosen as the center of the vortex core. Two crucial ingredients appear in this definition: (i) a real, rotation-invariant \emph{vortex profile function} $f(r)$ which goes to zero in the vortex core, and (ii) a vortex phase that winds $\ell$ times around the origin $i_0$ ($\ell$ being the \emph{vorticity}). These properties of a dissipative vortex naturally arise in an implementation with ultracold atoms based on optical vortex imprinting (see SI). They are fully analogous to those of a vortex in a Hamiltonian scenario. In particular, a $\pi$-flux is optically imprinted onto the matter system in the case $\ell =1$.

In what follows, we show that an \emph{odd} dissipative vortex (i.e. with odd vorticity) \emph{generically} traps a single, isolated MZM despite the seemingly trivial topological nature of the bulk. We focus on the simplest case of a single vortex with $\ell = 1$ and proceed in two main steps. First, we demonstrate that the vortex phase crucially modifies the topology of the system and construct a mapping to an effective 1D model. Second, we examine the effect of the dissipative ``boundary'' conditions imposed in the vortex core by the profile function $f(r)$, and show that the latter are responsible for the existence of a single MZM in the core.

\begin{figure}[b]
	\begin{center}
		\includegraphics[width=.98\columnwidth]{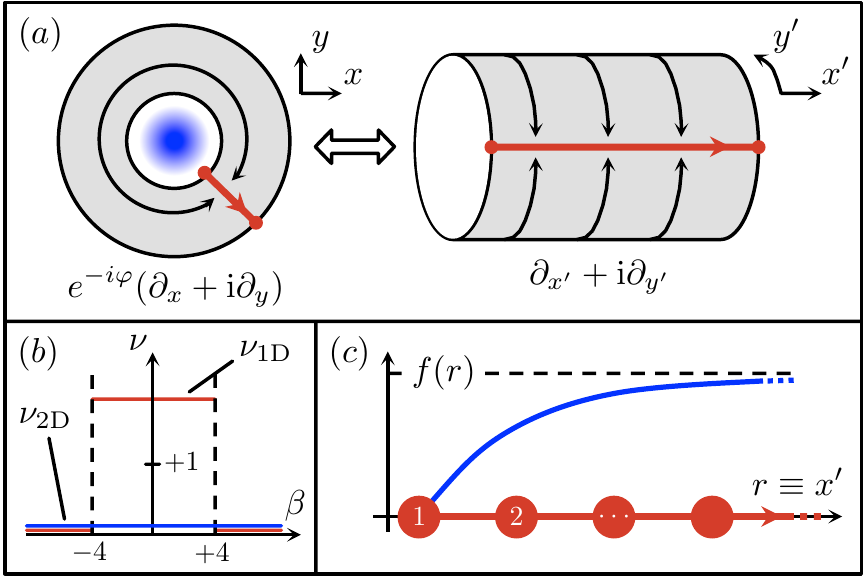}
		\caption{(Color online). Mechanism ensuring the existence of a \emph{single} MZM in the core of an odd dissipative vortex. (a), Left: $\ell = 1$ vortex on the plane, characterized by a core (blue) and a phase factor premultiplying the $p$-wave operator $\partial_x + \mathrm{i} \partial_y$. Right: Mapping from the (grey) annular region around the vortex core to the cylinder --where the vortex phase disappears from the relevant $p$-wave operator-- and reduction to a (chiral) 1D wire problem (red). (b) Topological phase diagram for the planar model of Eq.~\eqref{eqn:crossLindbladOperators} with no vortex (blue, characterized by the Chern number $\nu_{\text{2D}}$) and with a single $\ell = 1$ vortex (red, characterized by the winding number $\nu_{\text{1D}}$). (c) Dissipative boundary conditions of the 1D wire inherited from the vortex core profile (see text).}
		\label{fig:mechanism}
	\end{center}
\end{figure}

\emph{Mapping to a chiral 1D wire} -- To exemplify how an odd dissipative vortex changes the topology of the system, we consider the case $\ell = 1$ and assume, without loss of generality, that the vortex profile function $f(r)$ vanishes as $r \to 0$ and satisfies $f(r) = 1$ for $r > r_c$. We identify the region $r < r_c$ as the vortex core. In order to capture the generic properties of the bulk, we first focus on an annular region centered around the vortex core (Fig.~\ref{fig:mechanism}(a), left) where $f(r)$ is constant and the relevant $p$-wave operator carries the vortex phase, $e^{-\mathrm i \varphi} (\partial_x + \mathrm{i} \partial_y)$ (in the continuum limit, for simplicity). As made explicit in the SI, our model is, in this region, formally equivalent to a cylinder model with $p$-wave operator of the form $\partial_{x'} + \mathrm{i} \partial_{y'}$ ($(x', y')$ being Cartesian coordinates on the cylinder; see Fig.~\ref{fig:mechanism}(a), right) as in the original translation-invariant model on the plane (see Eq.~\eqref{eqn:crossLindbladOperators}). There exists therefore a one-to-one correspondence between the original planar model with an $\ell = 1$ vortex and the same model in cylinder geometry with no vortex. Physically, this stems from the fact that the gauge field $e^{-i \varphi}$ imposed on the plane to describe the vortex naturally arises on the cylinder owing to the extrinsic curvature of the latter.

In order to further extract the essence of the single-vortex problem, we revert to the original lattice description and take advantage of the translation invariance along the $y'$-direction. Defining Fourier-transformed Lindblad operators $L_{x'} (k_{y'}) \propto \sum_{y'} e^{-\mathrm{i} k_{y'} y'} L_{x', y'}$, we reduce the system to a stack of 1D wires that can be investigated along the lines of Ref.~\cite{Diehl11}. In the two momentum sectors corresponding to $k_{y'} = 0$ and $\pi$, the relevant 1D Lindblad operators take the form
\begin{align} \label{eqn:1DCrossLindbladOperators}
	L_i = \beta' \, a^\dagger_i & + ( a^\dagger_{i-1} + a^\dagger_{i+1} ) + ( -a_{i-1} + a_{i+1} ),
\end{align}
where $i$ indexes the lattice sites in the $x'$-direction (such that $L_i \equiv L_{x'} (k_{y'} = 0 \text{ or } \pi)$) and $\beta' \equiv \beta + 2$ $(\beta - 2)$ for $k_{y'} = 0$ ($\pi$). In these two particular sectors, the system thus reduces to a 1D wire with chiral symmetry~\cite{Ryu10}. The steady-state bulk properties of this wire can be unveiled in the infinite-size limit, in which case the Lindblad operators form a complete set of anticommuting operators, leading to a pure steady state described by a real unit vector $\vect{n}_k$ as above (see Eq.~\eqref{chernNumber}). Owing to chiral symmetry, this state can be characterized by a ``winding number'' topological invariant $\nu_{\text{1D}}$~\cite{Ryu02}. As detailed in the SI, we obtain
\begin{align} \label{eqn:windingNumber}
	\nu_{\text{1D}} \equiv \tfrac{1}{2\pi} \int_{\mathrm{BZ}} dk \; \vect{a} \cdot ( \vect{n}_k \times \partial_k \vect{n}_k ) = 2
\end{align}
for $\abs{\beta'} < 2$ (i.e. for $0 < \abs{\beta} < 4$), and $\nu_{\text{1D}} = 0$ otherwise ($\vect{a}$ being a unit vector orthogonal to $\vect{n}_k$ whose existence is guaranteed by chiral symmetry). We thus find non-trivial topological order in the parameter range delimited by the special points $\beta = 0$ and $\pm 4$ where the Chern number $\nu_{\text{2D}}$ exhibits discontinuities. Fig.~\ref{fig:mechanism}(b) illustrates the resulting topological phase diagram: non-equilibrium topological quantum phase transitions occur at the critical values $\beta = \pm 4$, while $\beta = 0$ corresponds to an isolated point separating two topologically equivalent phases. In agreement with the full 2D model, the dissipative gap of the 1D wire closes at each of these values (see SI).

\begin{figure}[t]
	\begin{center}
		\includegraphics[width=\columnwidth]{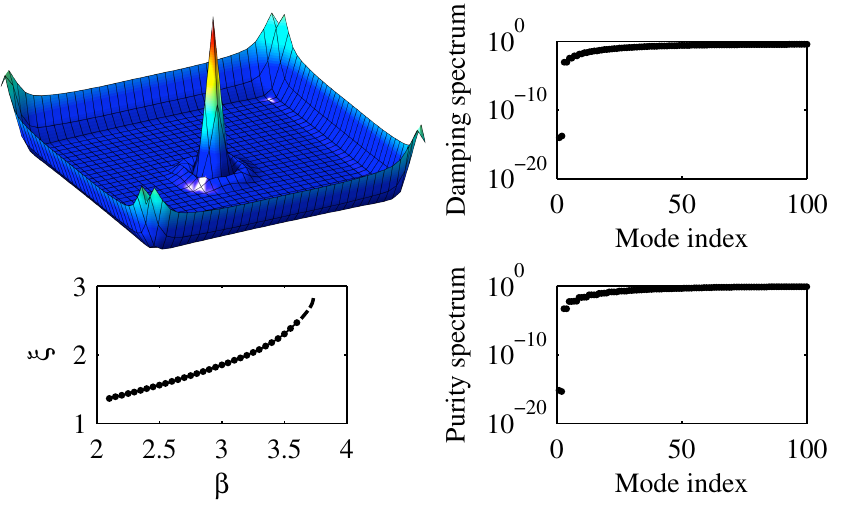}
		\caption{(Color online). Numerical results for a single dissipative vortex with $\ell = 1$ on a square lattice of $35 \times 35$ sites with unit spacing and $\kappa = 1$. Left: Typical form of the MZMs localized in the vortex core and on the edge, respectively (here for $\beta = 3$), and localization length scale $\xi$ associated with the MZM trapped in the core. Right: Low-lying part of the damping and purity spectra for $\beta = 2$, both featuring a gap with two zero eigenvalues. All results were obtained for a vortex profile $f(r) = (r/d) e^{-(r/\xi)^2}$ with $d = 10$ and $\xi = 20$.}
		\label{fig:numericalResults}
	\end{center}
\end{figure}

\emph{Dissipative boundary conditions} -- The fact that we can identify a non-trivial topological invariant in an effective 1D model describing the region surrounding the vortex core strongly supports the existence of interesting ``edge'' physics inside the latter. The bulk steady state of the 1D wire, which is pure, can equivalently be described as the ground state of the parent Hamiltonian $H_\text{parent} = \sum_i L_i^\dagger L_i$. Bulk-edge correspondence arguments based on $H_\text{parent}$ would therefore suggest the existence of $\nu_{\text{1D}} = 2$ MZMs at the edges of the wire --i.e., in particular, in the vortex core. In our general dissipative setting, however, bulk-edge correspondence arguments can only be formulated in the presence of a dissipative gap \emph{and} in the absence of modes corresponding to subspaces in which the steady state is completely mixed, which we refer to as \emph{purity zero modes} (see SI). Keeping in mind that the presence of \emph{dissipative} boundary conditions can potentially give rise to such modes, we now proceed to examine the ``edge'' physics that emerges in the vortex core as a result of the dissipative boundary conditions imposed by the vortex profile function $f(r)$. To this end, we first extend the above mapping by reducing the inner radius of the annular region shown in Fig.~\ref{fig:mechanism}(a) to zero. The resulting extended 1D wire is depicted in Fig.~\ref{fig:mechanism}(c); its first, leftmost site corresponds to the center of the vortex core where $f(r)$ vanishes, as shown in blue. The fact that $f(r)$ varies from site to site in the vortex core crucially leads to a violation of the purity condition $\{ L_i, L_j \} = 0$ (for all $i, j$) which is satisfied in the bulk, as mentioned above. One can easily verify that the anticommutator $\{ L_i, L_j \}$ increasingly deviates from zero for $| i - j | \leq 2$ upon approaching the left edge of the wire. As a consequence, the steady state increasingly loses purity and departs from the ground state of $H_\text{parent}$ featuring a pair of MZMs. Remarkably, \emph{one and only one} MZM survives at the edge of the wire --or, equivalently, in the vortex core-- in the full parameter range $0 < \abs{\beta} < 4$ associated with a non-trivial topological phase. This mode is explicitly constructed in the SI, and is shown to be exponentially localized, on a characteristic length scale $\xi = \xi(\abs{\beta'}) \sim 1/\abs{\log{(\abs{\beta'}/2)}}$ which diverges at $\abs{\beta'} = 2$, i.e. at $\beta = \pm 4$. At these values which coincide with the critical points found above, the norm of the wavefunction associated with the MZM diverges with the length of the wire, while the dissipative gap closes in the bulk. This further confirms the onset of a non-equilibrium topological quantum phase transition. The second MZM naively expected in the vortex core, by contrast, acquires a finite damping rate and becomes an environmental degree of freedom with no correlations with the rest of the system. As shown in the SI, any zero mode of $H_\text{parent}$ which acquires a finite damping rate is effectively traced out of the system in steady state, \emph{independently} of the initial conditions. We refer to such a mode with no Hamiltonian counterpart as an \emph{intrinsic} purity zero mode, in accordance with the fact that its existence is intrinsic to the dissipative dynamics itself.

The above findings are supported by extensive numerical simulations, confirming the existence of a single MZM trapped in the vortex core in the full parameter range $0 < \abs{\beta} < 4$ for arbitrary vortex profile functions, as well as the divergence of the corresponding localization length upon approaching the critical points $\beta = \pm 4$ (see Fig.~\ref{fig:numericalResults}). As expected, an intrinsic purity zero mode is found in the vortex core. Similar features are obtained for odd vortices with $\ell > 1$; even vortices, in contrast, do not exhibit any MZM, as expected from the fact that the vortex phase can be gauged away in that case (see e.g. Ref.~\cite{Read00}). Our numerical results generally support the main conclusion that odd dissipative vortices generically trap single, isolated MZMs despite the seemingly trivial topological nature of the system. Following the arguments of Ref.~\cite{Diehl11}, we expect such vortices to exhibit non-Abelian exchange statistics when braided around each other through adiabatic parameter changes.


The authors would like to thank A. Akhmerov, M. Fleischhauer, S. Habraken, V. Gurarie and V. Lahtinen for insightful discussions. This work was supported by NCCR Quantum Science and Technology (NCCR QSIT), research instrument of the Swiss National Science Foundation (SNSF), an ERC Advanced Investigator Grant (A. I.), the Austrian Science Fund (FWF) through SFB FOQUS F4016-N16 and the START grant Y 581-N16 (S. D.), the European Commission (AQUTE, NAMEQUAM), the Institut f\"ur Quanteninformation GmbH and the DARPA OLE program.



\begin{thebibliography}{99}
\bibitem{Footnote1}
Time-reversal symmetry is preserved if and only if there exists a unitary matrix $\Theta$ such that $Q_\bk \equiv \mathbf{n}_\bk \boldsymbol{\sigma}$ satisfies $\Theta^\dagger Q_{-\bk} \Theta = Q_\bk$. No such matrix exists in our case.
\bibitem{Footnote2}
Since time-reversal symmetry is broken, the vanishing of the Chern number is not a priori guaranteed (unlike in Ref.~\cite{Fu08} for example).
\end{thebibliography}
\end{document}



\title{Supplementary Information for ``Majorana Modes in Driven-Dissipative Atomic Superfluids With Zero Chern Number''}

\author{C.-E. Bardyn}
\affiliation{Institute for Quantum Electronics, ETH Zurich, 8093 Zurich, Switzerland}

\author{M. A. Baranov}
\affiliation{Institute for Quantum Optics and Quantum Information of the Austrian Academy of Sciences, A-6020 Innsbruck, Austria}
\affiliation{Institute for Theoretical Physics, University of Innsbruck, A-6020 Innsbruck, Austria}
\affiliation{NRC ``Kurchatov Institute'', Kurchatov Square 1, 123182 Moscow, Russia}

\author{E. Rico}
\affiliation{Institute for Quantum Optics and Quantum Information of the Austrian Academy of Sciences, A-6020 Innsbruck, Austria}
\affiliation{Institute for Theoretical Physics, University of Innsbruck, A-6020 Innsbruck, Austria}

\author{A. \.Imamo\u glu}
\affiliation{Institute for Quantum Electronics, ETH Zurich, 8093 Zurich, Switzerland}

\author{P. Zoller}
\affiliation{Institute for Quantum Optics and Quantum Information of the Austrian Academy of Sciences, A-6020 Innsbruck, Austria}
\affiliation{Institute for Theoretical Physics, University of Innsbruck, A-6020 Innsbruck, Austria}

\author{S. Diehl}
\affiliation{Institute for Quantum Optics and Quantum Information of the Austrian Academy of Sciences, A-6020 Innsbruck, Austria}
\affiliation{Institute for Theoretical Physics, University of Innsbruck, A-6020 Innsbruck, Austria}

\maketitle


\section{Dissipative framework}

In the following section, we discuss the properties of the dissipative dynamics considered in the main text and its relation to (dissipative) Majorana modes. We start by briefly reviewing a mathematical framework originally introduced in Refs.~\cite{Prosen08, Eisert10} which provides an alternative to the second-quantized formalism used in the main text. In this description --made possible by the quadratic nature of the dynamics-- the density matrix is replaced by a \emph{correlation matrix} $\Gamma$ which encodes the single-particle correlations and fully describes the system when Gaussian initial states are assumed. The information contained in the Lindblad operators is encoded in a \emph{damping matrix} $X$ and a \emph{fluctuation matrix} $Y$. As we will show, the matrix $Y$ shares the formal properties of a Hamiltonian in a first-quantized (matrix) representation~\cite{Altland97}. The steady state, however, is determined by the combined properties of $X$ and $Y$. Dissipative Majorana zero modes, in particular, can be identified as zero modes of $X$. We will give precise conditions for the existence of such modes.

A crucial difference between the ground state of a Hamiltonian and the steady state of a dissipative dynamics lies in the purity of the state. A proper assessment of purity can be made by examining the \emph{purity spectrum}, i.e. the spectrum of the matrix $\Gamma^2$; pure states are signaled by a flat (degenerate) purity spectrum with eigenvalues equal to $-1$, while completely mixed subspaces are indicated by zero eigenvalues which can be traced to the existence of \emph{purity zero modes}. In the Hamiltonian ground-state setting, the state is pure by construction, up to a possible degenerate subspace whose purity is not determined by the dynamics itself. Such a degenerate subspace is found for example in the presence of Majorana zero modes. In that case, starting from an initial high-temperature state, the purity of the Majorana subspace generally vanishes. Since the mixedness is extrinsic to the dynamics, and rather results from specific initial conditions, we refer to the Majorana zero modes as \emph{extrinsic purity zero modes}. The dissipative analog of this situation is given by dissipative Majorana zero modes. Using the quantum optics terminology, such modes can be identified as giving rise to a \emph{decoherence-free subspace}, i.e. to a subspace that is decoupled from dissipation. In the dissipative setting, however, a different situation can be encountered: the purity of the Majorana subspace can be pinned to zero by virtue of the dissipative dynamics, independently of the initial conditions. In that case, we refer to the corresponding Majorana zero modes as \emph{intrinsic purity zero modes}. Below we establish criteria for the occurrence of this phenomenon with no Hamiltonian counterpart. We also derive useful necessary and sufficient conditions for pure steady states.

\subsection{Majorana zero modes and associated decoherence-free subspace}

We recall the form of the dissipative dynamics considered in the main text, namely,
\begin{align} \label{eqn:masterEquation}
	\partial_t \rho = \kappa \sum_{i = 1}^n \left( L_i \rho L_i^\dagger - \tfrac{1}{2} \{ L^\dagger_i  L_i, \rho \} \right),
\end{align}
with Lindblad operators $L_i$ that are linear combinations of fermionic operators $a_j^\dagger, a_j$ ($j = 1, \ldots, N$). As shown in Refs.~\cite{Prosen08, Eisert10}, the fact that Eq.~\eqref{eqn:masterEquation} is quadratic allows for a convenient simplification: introducing a basis of Majorana operators $c_{2i-1} \equiv a^\dagger_i + a_i$, $c_{2i} \equiv \mathrm{i} (a^\dagger_i - a_i)$ and assuming Gaussian initial states, one can focus exclusively on the time evolution of the real antisymmetric \emph{correlation matrix} $\Gamma_{ij} \equiv \tfrac{\mathrm{i}}{2} \operatorname{tr} ([ c_i, c_j ] \rho)$ which fully describes the state of the system. Expressing the Lindblad operators in the vector form $L_i = \vect{l}_{i}^T \vect{c}$, where $\vect{c}^T = (c_1, c_2, \ldots, c_{2N})$, one can recast Eq.~\eqref{eqn:masterEquation} as the following matrix equation:
\begin{align} \label{eqn:masterEquationMatrixForm}
	\partial_t \Gamma = -\{ X,\Gamma \} + Y,
\end{align}
where $X \equiv 2 \operatorname{Re} M$ and $Y \equiv 4 \operatorname{Im} M$, with $M \equiv \sum_i \vect{l}_{i} \otimes \vect{l}^\dagger_i$ (note that we have set $\kappa = 1$ in Eq.~\eqref{eqn:masterEquation}; this factor can be absorbed in the definition of $L_i$, without loss of generality). Physically, the real symmetric matrix $X$ describes the damping; the real antisymmetric matrix $Y$, on the other hand, describes the fluctuations in steady state $\tilde{\Gamma}$ through the steady-state equation $\{ X, \tilde{\Gamma} \} = Y$. We thus refer to $X$ and $Y$ as the \emph{damping} and \emph{fluctuation} matrices, respectively. Their explicit form reads
\begin{align} \label{eqn:XYMatrices}
	\begin{array}{lll}
		X & = & \sum_{i = 1}^{n} \left( \vect{l}_i \otimes \vect{l}^\dagger_i + \vect{l}^*_i \otimes \vect{l}^T_i \right), \\
		Y & = -2i & \sum_{i = 1}^{n} \left( \vect{l}_i \otimes \vect{l}^\dagger_i - \vect{l}^*_i \otimes \vect{l}^T_i \right).
	\end{array}
\end{align}
The damping matrix is real, symmetric, and positive semidefinite (by construction). It can therefore be spectrally decomposed as $X = \sum_{j = 1}^{2N} \lambda_j (\vect{v}_j \otimes \vect{v}^T_j)$ with eigenvalues $\lambda_j \in \mathbb{R}_{\geq 0}$ and associated eigenvectors $\vect{v}_j \in \mathbb{R}^{2N}$. The real eigenvectors $\vect{v}_j$ form an orthonormal basis and define a natural basis of Majorana operators $\gamma_j \equiv \vect{v}^T_j \vect{c} = \gamma^\dagger_j$ ($j = 1, 2, \dots, 2N$) associated with $X$ (we remark that this Majorana basis is not necessarily local, as opposed to the original basis given by $\vect{c}^T = (c_1, c_2, \ldots, c_{2N})$). The fact that they are real implies that $\vect{l}^\dagger_i \vect{v}_j = (\vect{l}^T_i \vect{v}_j)^*$ for all $j$, such that
\begin{align} \label{eqn:XMatrixElements}
	X_{jk} & \equiv \vect{v}^T_j X \vect{v}_k = 2 \sum_{i = 1}^{n} \operatorname{Re} {\{ (\vect{l}^T_i \vect{v}_j) (\vect{l}^T_i \vect{v}_k)^* \}} = \lambda_j \, \delta_{jk} \\
	\intertext{with $\lambda_j  = 2 \sum_{i = 1}^{N} \abs{\vect{l}^T_i \vect{v}_j}^2$, and} \label{eqn:YMatrixElements}
	Y_{jk} & \equiv \vect{v}^T_j Y \vect{v}_k = 4 \sum_{i = 1}^{n} \operatorname{Im} {\{ (\vect{l}^T_i \vect{v}_j) (\vect{l}^T_i \vect{v}_k)^* \}}.
\end{align}
Physically, the eigenvalues $\lambda_j$ correspond to the damping rates associated with the Majorana modes $\gamma_j$. As shown above, they are solely determined by the amount of overlap $\vect{l}^T_i \vect{v}_j$ between the eigenvectors $\vect{v}_j$ of $X$ and the vectors $\vect{l}_i$ representing the Lindblad operators in the original (local) Majorana basis. Since $\vect{l}^T_i \vect{v}_j = \{L_i, \gamma_j\}$, they can alternatively be seen as determined by the amount by which the Majorana modes $\gamma_j$ anticommute with the Lindblad operators $L_i$. In general, we will refer to a Majorana mode $\gamma_j$ as a Majorana \emph{zero} mode whenever the associated damping rate $\lambda_j$ vanishes. In light of the above discussion, a unit vector $\vect{v} \in \mathbb{R}^{2N}$ defines a Majorana zero mode $\gamma = \vect{v}^T \vect{c}$ if and only if the following equivalent conditions are satisfied:
\begin{enumerate}[\hspace{0.5cm}(i)]
	\item $X \vect{v} = 0$;
	\item $\vect{l}^T_i \vect{v} = 0$ for all $i \in \{1, 2, \dots, n\}$;
	\item $\{L_i, \gamma\} = 0$ for all $i \in \{1, 2, \dots, n\}$.
\end{enumerate}
Remembering the explicit form of the matrix $Y$, these conditions also imply $Y \vect{v} = 0$ (the converse is generally not true).

The existence of $m \leq 2N$ Majorana zero modes generally implies the existence of a $(m - m \, \text{mod} \, 2)$-dimensional subspace $V_0$ that is left invariant under the dissipative dynamics of Eq.~\eqref{eqn:masterEquationMatrixForm}. This stems from the fact that the state space of the system is isomorphic to the space of all real $2N \times 2N$ \emph{antisymmetric} correlation matrices $\Gamma$ (assuming Gaussian states), and is therefore even-dimensional (with dimension $2N(2N - 1)/2$), as must be $V_0$. As a consequence, Majorana modes can only decouple from the dissipative dynamics \emph{in pairs}. In the original Hilbert-space description of the system, $V_0$ translates as a \emph{decoherence-free subspace}~\cite{Lidar98} of dimension $2^{(m - m \, \text{mod} \, 2)}$.

\subsection{Purity zero modes and robustness of the Majorana zero modes}

In the matrix representation defined above, the purity of a state with correlation matrix $\Gamma$ can be inferred from the spectrum of $\Gamma^2$, which we refer to as the \emph{purity spectrum}. One can easily convince oneself that $\Gamma$ describes a pure state if and only if $\Gamma^2 = -1$, i.e. iff the purity spectrum is flat with all eigenvalues equal to $-1$. In particular, a mode $\tilde{\gamma} \equiv \vect{w}^T \vect{c}$ corresponding to an eigenvector $\vect{w}$ of $\Gamma^2$ with eigenvalue zero can be identified as a \emph{purity zero mode}, i.e. as a mode associated with a subspace in which the state is completely mixed. We distinguish two types of such modes: (i) the \emph{intrinsic} purity zero modes, which appear in steady state \emph{independently} of the initial conditions, and (ii) the \emph{extrinsic} purity zero modes, which solely result from mixed initial conditions and thus disappear if the system is prepared in a pure initial state. (Note, however, that a completely mixed initial state is a rather generic assumption when a spatially separated pair of Majorana modes is present.)

Keeping the above considerations in mind, we now proceed to examine the effect of dissipative perturbations on Majorana zero modes. We consider the general case where the dissipative dynamics gives rise to $m < 2N$ Majorana zero modes $\gamma_\alpha = \vect{v}^T_\alpha \vect{c}$, and define $V'_0 \subset \mathbb{R}^{2N}$ as the $m$-dimensional subspace associated with these modes (with Greek indices denoting quantities pertaining to the latter). In addition, we assume that (Gaussian) correlations are initially present between the Majorana zero modes, as determined by the initial correlation matrix $\Gamma_{\alpha \beta}(t = 0) = \tfrac{\mathrm{i}}{2} \operatorname{tr} ([ \gamma_\alpha, \gamma_\beta ] \rho(t = 0))$. By definition, the dissipative dynamics restricted to $V'_0$ is trivial, namely, $\partial_t \Gamma_{\alpha \beta} = 0$ for all pairs of indices $\alpha, \beta \in \{ 1, 2, \ldots, m \}$. In order to examine under what conditions the evolution remains trivial in $V'_0$ when the system is perturbed, we extend the dissipative dynamics by introducing an additional Lindblad operator $L_{n+1} = \vect{l}^T_{n+1} \vect{c}$. The $X$ and $Y$ matrices restricted to $V'_0$ --which were identically zero-- then become
\begin{align}
	X_{\alpha \beta} \equiv \vect{v}^T_\alpha X \vect{v}_\beta & = 2 \, \operatorname{Re} {\{ (\vect{l}^T_{n+1} \vect{v}_\alpha)(\vect{l}^T_{n+1} \vect{v}_\beta)^* \}}, \\
	Y_{\alpha \beta} \equiv \vect{v}^T_\alpha Y \vect{v}_\beta & = 4 \, \operatorname{Im} {\{ (\vect{l}^T_{n+1} \vect{v}_\alpha) (\vect{l}^T_{n+1} \vect{v}_\beta)^* \}}.
\end{align}
Assuming that $L_{n+1}$ acts quasi-locally and that the Majorana zero modes are localized and spaced sufficiently far away from each other (such that $\{ \gamma_\alpha, \gamma_\beta \} = v_\alpha^T v_\beta = 0$ up to exponentially small corrections), two cases can occur: either (i) $\vect{l}^T_{n+1} \vect{v}_\alpha = \{ L_{n+1}, \gamma_\alpha \} = 0$ for all $\alpha \in \{ 1, 2, \dots, m \}$, or (ii) $\vect{l}^T_{n+1} \vect{v}_\mu = \{ L_{n+1}, \gamma_\mu \} \neq 0$ for a single $\mu \in \{ 1, 2, \dots, m \}$. Clearly, the dissipative dynamics restricted to $V'_0$ remains trivial in the first case: the Majorana zero modes survive, and all correlations initially present in $V'_0$ are preserved over arbitrary long times. In the second case, however, the situation changes: although the off-diagonal elements of $X$ and $Y$ remain trivial owing to the local nature of $L_{n+1}$ and $\gamma_\alpha$ --such that $Y$ remains identically zero-- the diagonal element $X_{\mu \mu}$ of the symmetric matrix $X$ becomes non-zero. More specifically, the Majorana mode $\gamma_\mu$ acquires a non-zero damping rate $\lambda_\mu = 2 \abs{\vect{l}^T_{n+1} \vect{v}_\mu}^2$ (independent of the system size), and the dissipative dynamics reads $\partial_t \Gamma_{\alpha \mu} = -\lambda_\mu \Gamma_{\alpha \mu}$ for all $\alpha \in \{ 1, 2, \dots, m \}$. In other words, the introduction of an additional quasi-local Lindblad operator $L_{n+1}$ which anticommutes with all Majorana zero modes but $\gamma_\mu$ has the effect of destroying all non-local correlations initially present in $V'_0$ that involve the latter. After a time $t \sim 1/\lambda_\mu$, $\gamma_\mu$ is effectively traced out of the system and emerges as an \emph{intrinsic} purity zero mode: for arbitrary initial states, a zero eigenvalue appears in the steady-state purity spectrum.

In summary, we have shown that quasi-local dissipative processes can generally remove Majorana zero modes from the system, thereby resulting in a loss of purity. A necessary and sufficient condition for a specific Majorana zero mode $\gamma$ to disappear under the effect of a particular dissipative process given by a Lindblad operator $L$ is given by $\{ L, \gamma \} \neq 0$. Although we have used the word ``perturbation'' above, we remark that this condition can be satisfied in a controlled way, by carefully \emph{engineering} dissipation. In fact, the model presented in the main text provides us with a paradigmatic example of such controlled dissipation: due to the generic form of the engineered dissipative boundary conditions, the removal of a single Majorana zero mode naturally occurs (see main text).

\subsection{Hamiltonian ground states vs. Liouvillian steady states}

The parent Hamiltonian $H_{\text{parent}} \equiv \sum_i L^\dagger_i L_i$ introduced in the main text plays a crucial role in understanding the similarities and differences between Hamiltonian and Liouvillian dynamics. Using the Majorana basis defined above, one can write $H_{\text{parent}} =  \vect{c}^T (\sum _i \vect{l}^*_i \otimes \vect{l}_i^T) \vect{c} = \tfrac{\mathrm{i}}{4} \vect{c}^T Y \vect{c}$; the fluctuation matrix $Y$ thus clearly appears as the counterpart of the parent Hamiltonian in the matrix representation given by Eq.~\eqref{eqn:masterEquationMatrixForm}. This one-to-one correspondence between $H_{\text{parent}}$ and $Y$ clearly shows that the latter contains at least partial information about the dissipative dynamics. As we now proceed to prove, the information provided by $H_{\text{parent}}$ --or, equivalently, by $Y$-- is sufficient to fully describe the dissipative dynamics whenever the Lindblad operators satisfy $\{ L_i, L_j \} = 0$ for all $i, j$. This result will emerge as a by-product of the more general result proved below, which can be stated as follows:

\emph{Theorem}. For the purely dissipative dynamics defined by Eq.~\eqref{eqn:masterEquation}, the following statements are equivalent:
\begin{enumerate}[\hspace{0.5cm}(i)]
	\item There exists initial conditions leading to a pure steady state;
	\item The dissipative dynamics admits an equivalent Hamiltonian description, and the latter is provided by the parent Hamiltonian $H_{\text{parent}} \equiv \sum_i L^\dagger_i L_i$;
	\item The Lindblad operators form a set of anticommuting operators, i.e. $\{ L_i, L_j \} = 0$ for all $i,j$ $\in \{1, 2, \dots, n\}$;
	\item $[X, Y] = 0 \text{ and } X^2 = -\tfrac{1}{4} Y^2$.
\end{enumerate}
If there are as many non-zero Lindblad operators as fermionic sites --such that the Lindblad operators form a \emph{complete} set of anticommuting operators-- these conditions ensure that the steady state is pure and unique.

\emph{Proof}. Let us first assume that $\{ L_i, L_j \} = 0$ is satisfied for all $i, j$. In the matrix representation defined above, this condition translates as $\vect{l}_i^T \vect{l}_j = \vect{l}_i^\dagger \vect{l}^*_j = 0$ for all $i,j$. Remembering the explicit form of the $X$ and $Y$ matrices (see Eq.~\eqref{eqn:XYMatrices}), we then find
\begin{align} \label{eqn:XYrelation}
	[X, Y] = [X, \mathrm{i} Y] = 0 \text{ and } X^2 = -\tfrac{1}{4} Y^2 = (\tfrac{\mathrm{i}}{2} Y)^2,
\end{align}
as can easily be verified. The first equation above implies that the Hermitian matrices $X$ and $\mathrm{i} Y$ can be diagonalized simultaneously by a unitary transformation $U$. We denote the corresponding diagonal matrices as $D_X$ and $D_Y$, and notice that $D_Y \equiv \mathrm{i} U^\dagger Y U = \text{diag}(\lambda_{Y, 1}, -\lambda_{Y, 1}, \ldots, \lambda_{Y, N}, -\lambda_{Y, N})$ with $\lambda_{Y, j} \geq 0$ must be satisfied owing to the antisymmetry of $Y$. Using the second equation above, we then obtain $D_X^2 = \tfrac{1}{4} D_Y^2$, which in turn implies $D_X \equiv U^\dagger X U = \text{diag}(\lambda_{X, 1}, \lambda_{X, 1}, \ldots, \lambda_{X, N}, \lambda_{X, N})$ with $\lambda_{X, j} = \lambda_{Y, j}/2 \geq 0$ (remember that $X$ is positive semidefinite). The spectra of $X$ and $\mathrm{i} Y$ are thus in one-to-one correspondence (in particular, the spectrum of $X$ is doubly degenerate). Therefore, the fluctuation matrix $Y$ --or, equivalently, the parent Hamiltonian $H_{\text{parent}}$-- contains the same amount of information about the dissipative dynamics as the damping matrix $X$. In other words, the parent Hamiltonian provides a complete description of the dissipative dynamics when the Lindblad operators form a set of anticommuting operators.

As dictated by Eq.~\eqref{eqn:masterEquationMatrixForm}, the steady-state correlation matrix $\tilde{\Gamma}$ must satisfy the equation $\{ X, \tilde{\Gamma} \} = Y$, which can equivalently be written as $\{ D_X, \tilde{\Gamma}' \} = D_Y$ with $\tilde{\Gamma}' \equiv \mathrm{i} U^\dagger \tilde{\Gamma} U$. It should be clear that this steady-state equation does not determine the form of $\tilde{\Gamma}'$ in the decoherence-free subspace corresponding to $\lambda_{X, j} = 0$ (as argued previously, the latter solely depends on the initial conditions). However, one can easily verify that the steady-state equation constrains $\tilde{\Gamma}'$ to a block-diagonal form $\oplus_j \, \mathrm{i} \sigma^y$ in the subspace corresponding to $\lambda_{X, j} > 0$ owing to the tight relation between $D_X$ and $D_Y$ derived above. With proper initial conditions, it is therefore possible to obtain $\tilde{\Gamma}' = \oplus_{j = 1}^N \, \mathrm{i} \sigma^y$, which satisfies $\tilde{\Gamma}'^2 = \tilde{\Gamma}^2 = -1$. This concludes the first part of our proof.

Let us now show that the reverse is true, namely, that the existence of a pure steady state $\tilde{\Gamma}$ implies $\{ L_i, L_j \} = 0$ for all $i, j$. We first notice that $[ \tilde{\Gamma}, Y ] = [ \tilde{\Gamma}, \{ X, \tilde{\Gamma} \} ] = 0$ directly follows from the fact that $\tilde{\Gamma}$ satisfies $\{ X, \tilde{\Gamma} \} = Y$ and $\tilde{\Gamma}^2 = -1$. $\mathrm{i} \Gamma$ and $\mathrm{i} Y$ can thus be diagonalized simultaneously by a unitary transformation $U$, yielding $D_{\tilde{\Gamma}} \equiv \mathrm{i} U^\dagger \tilde{\Gamma} U = \text{diag}(1, -1, \ldots, 1, -1)$ and $D_Y \equiv \mathrm{i} U^\dagger Y U = \text{diag}(\lambda_{Y, 1}, -\lambda_{Y, 1}, \ldots, \lambda_{Y, N}, -\lambda_{Y, N})$ with $\lambda_{Y, j} \geq 0$. Expressing the steady-state equation in the basis defined by $U$, we then obtain $\{ U^\dagger X U, \Gamma' \} = Y'$, which in turn implies $U^\dagger X U = \text{diag}(\lambda_{X, 1}, \lambda_{X, 1}, \ldots, \lambda_{X, N}, \lambda_{X, N})$ with $\lambda_{X, j} = \lambda_{Y, j}/2 \geq 0$. Similarly as above, the spectrum of $X$ is therefore in one-to-one correspondence with the spectrum of $Y$; in particular, it is doubly degenerate. One can easily verify that $X^2 = -\tfrac{1}{4} Y^2$, as in Eq.~\eqref{eqn:XYrelation}. Remembering the explicit form of the $X$ and $Y$ matrices (see Eq.~\eqref{eqn:XYMatrices}), we thus obtain
\begin{align}
	\sum_{i,j} ( \vect{l}^*_i \otimes \vect{l}^T_i)( \vect{l}_j \otimes \vect{l}^\dagger_j) = \sum_{i,j} (\vect{l}^T_i \vect{l}_j )( \vect{l}^*_i \otimes \vect{l}^\dagger_j) = 0,
\end{align}
which, since $\vect{l}_i$ and $\vect{l}_j$ have support in different subspaces for $i \neq j$, can only be satisfied if $\{ L_i, L_j \} = \vect{l}^T_i \vect{l}_j = 0$ for all $i, j$. This concludes the second and last part of our proof. $\blacksquare$

The above discussion shows that the damping and fluctuation matrices $X$ and $Y$ are in one-to-one correspondence --and therefore \emph{both} have full information about the dissipative dynamics-- if and only if there exists a pure steady state or, equivalently, if and only if the Lindblad operators form a set of anticommuting operators. Remembering that the fluctuation matrix is the exact counterpart of the parent Hamiltonian (i.e. $H_{\text{parent}} = \tfrac{\mathrm{i}}{4} \vect{c}^T Y \vect{c}$), we conclude that $H_{\text{parent}}$ fully characterizes the dissipative dynamics whenever the steady state of the latter is pure. In that case, its ground state exactly coincides with the steady state, and its ground-state degeneracy --if any-- reflects the existence of a decoherence-free subspace associated with zero-damping Majorana modes $\gamma = \gamma^\dagger$ satisfying the orthogonality condition $\{ L_i, \gamma \} = 0$ for all $i$.

In the more general case where the steady state of Eq.~\eqref{eqn:masterEquation} is mixed, the one-to-one correspondence between $X$ and $Y$ breaks down and the parent Hamiltonian does not provide a complete description of the dissipative dynamics anymore. The existence of Majorana zero modes is then solely determined by the form of the damping matrix $X$. In particular, a zero mode of $H_{\text{parent}}$ (or $\mathrm{i} Y$) does not necessarily correspond to a zero mode of $X$, although the converse is always true (see discussion following Eqs.~\eqref{eqn:XMatrixElements} and~\eqref{eqn:YMatrixElements} above). Most crucially, any zero mode of $H_{\text{parent}}$ which does not coincide with a zero mode of $X$ is effectively traced out of the system in steady state, thereby giving rise to an \emph{intrinsic} purity zero mode. This effect lies at the very heart of the model constructed in the main text: in that model, $H_{\text{parent}}$ exhibits isolated \emph{pairs} of Majorana zero modes in vortices of odd vorticity or on the edge of the system. However, one member of each pair is traded for a purity zero mode as a result of dissipation, so that \emph{single} Majorana zero modes are obtained.

To conclude, we remark that the well-known Hamiltonian mechanisms of topological protection embedded in $H_{\text{parent}}$ do not extend to $X$. Indeed, while the antisymmetry of $Y$ ensures that the Majorana zero modes of $H_\text{parent}$ come in pairs and are thus unaffected by local perturbations when spatially separated, the symmetry of $X$ does not allow for such a mechanism. Majorana zero modes are therefore generally not protected against arbitrary local dissipative perturbations (as shown explicitly above). Although this effect peculiar to dissipation is potentially harmful, as has been recognized in Refs.~\cite{Chamon11, Trauzettel11} in another context, the model constructed in the main text shows that it can also lead to intriguing physics such as the transformation of Abelian vortices into non-Abelian ones. In general, engineered dissipation allows us to prepare topological phases with unexpected edge physics and no Hamiltonian counterpart.


\section{Calculation of the topological invariants and relation between 2D and 1D models}

In the following section, we give an explicit derivation of the Chern and winding numbers presented in the main text (see Eqs. (3) and (5)). We recall that both of these quantities are \emph{bulk} topological invariants, which are only defined in the limit of an infinite system. In what follows, we will therefore always assume the system to be infinite and translation invariant, for convenience. To conclude the section, we provide the main ingredient for mapping the 2D vortex model to a 1D model as argued in the main text.

\subsection{2D case: Chern number $\nu_{\text{2D}}$}

Using the notations of the main text, we consider the general case of a 2D system driven by momentum-space Lindblad operators of the form $L_\bk = u_\bk a_\bk + v_\bk a^\dagger_{-\bk}$, and define a ``Fermi surface'' $\mathcal{F}$ as the following set of points in the Brillouin zone (BZ):
\begin{align} \label{eqn:fermiSurface}
	\mathcal{F} = \{ \bk: \, \abs{\varphi_{\bk}} = 1 \},
\end{align}
where $\varphi_\bk = \abs{\varphi_{\bk}} e^{\mathrm{i} \theta_\bk}$ is the so-called \emph{pair wavefunction}, defined as $\varphi_\bk = v_\bk / u_\bk$. Since the Brillouin zone corresponds to a torus, this Fermi surface does not necessarily admit a well-defined interior and a natural orientation. Following Ref.~\cite{Cheng10}, we define an \emph{electron-like} region $\mathcal{E} = \{ \bk: \, \abs{\varphi_{\bk}} > 1 \}$ and a \emph{hole-like} region $\mathcal{H} =  \{ \bk: \, \abs{\varphi_{\bk}} < 1 \}$, and choose the Fermi surface to be oriented positively whenever the electron-like region is on the left as we go along it. With this convention, the interior of $\mathcal{F}$ can be identified with $\mathcal{E}$. In general, the electron-like region corresponds to disconnected (simply or multiply connected) regions of the Brillouin zone, and the Fermi surface $\mathcal{F}$ can be described as a finite union of piecewise smooth, simple closed curves $\mathcal{F}_\lambda$, i.e. $\mathcal{F} = \bigcup_\lambda \mathcal{F}_\lambda$ (more pathological Fermi surfaces can be encountered, but the Chern number can only be defined in that case). As shown in Ref.~\cite{Cheng10}, the Chern number can then be expressed as a sum of winding numbers $W_\lambda$ around each of the closed curves $\mathcal{F}_\lambda$, namely,
\begin{align} \label{eqn:chernNumber}
	\nu_{\text{2D}} & \equiv \frac{1}{4\pi} \int_{\mathrm{BZ}} d^2 \vect{k} \; \vect{n}_\bk \cdot ( \partial_{k_x} \vect{n}_\bk \times \partial_{k_y} \vect{n}_\bk ) \nonumber \\
	& = \sum_\lambda W_\lambda,
\end{align}
with winding numbers
\begin{align} \label{eqn:windingNumbers}
	W_\lambda & = \frac{1}{2\pi} \oint_{\mathcal{F}_\lambda} \nabla_\bk \theta_\bk \cdot d\bk \nonumber \\
	& = \frac{1}{2\pi} \oint_{\mathcal{F}_\lambda} (\partial_{k_x} \theta_\bk dk_x + \partial_{k_y} \theta_\bk dk_y).
\end{align}
If $\partial_{k_x} \partial_{k_y} \theta_\bk$ and $\partial_{k_y} \partial_{k_x} \theta_\bk$ are continuous functions on the electron-like region enclosed by $\mathcal{F}_\lambda$, Green's theorem can be applied and gives
\begin{align}
	W_\lambda = \frac{1}{2\pi} \iint_{\text{int} \mathcal{F}_\lambda} \left( \partial_{k_x} \partial_{k_y} \theta_\bk - \partial_{k_y} \partial_{k_x} \theta_\bk \right) dk_x dk_y = 0.
\end{align}
In other words, $W_\lambda$ trivially vanishes. In order to allow for non-trivial contributions to the Chern number, it is therefore necessary to have discontinuities in the second-order mixed derivatives of $\theta_\bk$ in the electron-like region (as a matter of fact, in the absence of such discontinuities, one can always perform a gauge transformation to get rid of the phase $\theta_\bk$). We note that a necessary and sufficient condition for a particular discontinuity point $\bk_i$ to be contained in $\mathcal{E}$ is given by
\begin{align} \label{eqn:limit}
	\lim_{{\bf{k}} \to {\bf{k}}_i} \left( |\varphi_{\bf{k}} | - 1 \right) > 0.
\end{align}
When this condition is satisfied, $\bk_i$ contributes to the Chern number by an integer value $w_i$ equal to the number of times that the phase $\theta_\bk$ winds by $2 \pi$ when going around $\bk_i$ (in the counter-clockwise direction), as dictated by Eq.~\eqref{eqn:windingNumbers}. This provides us with a simple way of determining the Chern number.

Let us now examine the specific case considered in the main text, with Lindblad operators defined as
\begin{align} \label{eqn:crossLindbladOperators}
	\begin{split}
		L_i = \beta \, a^\dagger_i \; + \; & ( a^\dagger_{i + \bf{e}_x} + \hphantom{\mathrm{i}} a^\dagger_{i + \bf{e}_y} + \phantom{\mathrm{i}} a^\dagger_{i - \bf{e}_x} + \hphantom{\mathrm{i}} a^\dagger_{i - \bf{e}_y} ) \\
		\; + \; & ( a_{i + \bf{e}_x} + \mathrm{i} a_{i + \bf{e}_y} - \hphantom{\mathrm{i}} a_{i - \bf{e}_x} - \mathrm{i} a_{i - \bf{e}_y} ),
	\end{split}
\end{align}
where we have set $\alpha = 1$ and $\phi = 0$ (see main text), without loss of generality. Expressed in momentum space, these operators take the form
\begin{align} \label{eqn:Lops}
	L_\bk & = u_\bk a_\bk + v_\bk a^\dagger_{-\bk}, \nonumber \\
	u_\bk & = 2 \mathrm{i} ( \sin{(k_x)} + \mathrm{i} \sin{(k_y)} ), \nonumber \\
	v_\bk & = \beta + 2 ( \cos{(k_x)} + \cos{(k_y)} ).
\end{align}
The minimum of their ``norm''
\begin{align}
	\mathcal{N}_\bk \equiv \{ L^\dagger_\bk, L_\bk \} = \abs{u_\bk}^2 + \abs{v_\bk}^2
\end{align}
can easily be seen to be proportional to the dissipative gap. Since it vanishes for $\beta = -4$ at $(k_x, k_y) = (0, 0)$, $\beta = 0$ at $(k_x, k_y) = (\pi, 0), (0, \pi)$, and $\beta = 4$ at $(k_x, k_y) = (\pi, \pi)$, we anticipate the existence of four topological phases, depending on the value of $\beta$. As shown above, the Chern number is solely determined by the existence of discontinuity points and the behavior of the pair wavefunction $\varphi_\bk$ around them. Here we have
\begin{align}
	\varphi_\bk = \frac{\beta + 2 ( \cos{(k_x)} + \cos{(k_y)} )}{2 \mathrm{i} ( \sin{(k_x)} + \mathrm{i} \sin{(k_y)} )},
\end{align}
and the phase $\theta_\bk$ is defined everywhere except at the four discontinuity points $\bk_i \in \{ (0, 0), (0, \pi), (\pi, 0), (\pi, \pi) \}$ which coincide with the gap-closing points found above (note that these points are invariant under time-reversal symmetry $\bk_i \rightarrow -\bk_i$). The behavior of $\varphi_\bk$ around them is given by
\begin{align} \label{eqn:behaviorAroundDiscontinuityPoints}
	\lim_{\bk \to (0, 0)} \varphi_{\bk} & = \lim_{\epsilon \to 0} \frac{  \beta + 2 \left[ \cos{\left( \epsilon \cos{\theta} \right) }  + \cos{\left( \epsilon \sin{\theta} \right) }  \right]   }{ 2 \mathrm{i} \left[ \sin{\left( \epsilon \cos{\theta} \right) } + \mathrm{i} \sin{\left( \epsilon \sin{\theta} \right) } \right]  } \nonumber \\
	& = \lim_{\epsilon \to 0} \frac{\beta + 4}{2 \mathrm{i} \epsilon} e^{- \mathrm{i} \theta} \nonumber \\
	\lim_{\bk \to (0, \pi)} \varphi_{\bk} & = \lim_{\epsilon \to 0} \frac{  \beta + 2 \left[ \cos{\left( \epsilon \cos{\theta} \right) }  + \cos{\left( \pi+ \epsilon \sin{\theta} \right) }  \right]   }{ 2 \mathrm{i} \left[ \sin{\left( \epsilon \cos{\theta} \right) } + \mathrm{i} \sin{\left( \pi+\epsilon \sin{\theta} \right) } \right]  } \nonumber \\
	& = \lim_{\epsilon \to 0} \frac{\beta}{2 \mathrm{i} \epsilon} e^{\mathrm{i} \theta} \nonumber \\
	\lim_{\bk \to (\pi, 0)} \varphi_{\bk} & = \lim_{\epsilon \to 0} \frac{  \beta + 2 \left[ \cos{\left( \pi + \epsilon \cos{\theta} \right) }  + \cos{\left(  \epsilon \sin{\theta} \right) }  \right]   }{ 2 \mathrm{i} \left[ \sin{\left(\pi + \epsilon \cos{\theta} \right) } + \mathrm{i} \sin{\left( \epsilon \sin{\theta} \right) } \right]  } \nonumber \\
	& = \lim_{\epsilon \to 0} \frac{\beta}{-2 \mathrm{i} \epsilon} e^{\mathrm{i} \theta} \nonumber \\
	\lim_{\bk \to (\pi, \pi)} \varphi_{\bk} & = \lim_{\epsilon \to 0} \frac{  \beta + 2 \left[ \cos{\left( \pi + \epsilon \cos{\theta} \right) }  + \cos{\left( \pi + \epsilon \sin{\theta} \right) }  \right]   }{ 2 \mathrm{i} \left[ \sin{\left(\pi + \epsilon \cos{\theta} \right) } + \mathrm{i} \sin{\left(\pi + \epsilon \sin{\theta} \right) } \right]  } \nonumber \\
	& = \lim_{\epsilon \to 0} \frac{\beta - 4}{- 2 \mathrm{i} \epsilon} e^{- \mathrm{i} \theta},
\end{align}
where $\epsilon$, $\theta$ are polar coordinates defined around the point of interest. The phase $\theta_\bk$ is thus found to rotate in the counter-clockwise direction around the discontinuity points $\bk_1 = (0, 0)$ and $\bk_2 = (\pi, \pi)$, and in the clockwise direction around $\bk_3 = (0, \pi)$ and $\bk_4 = (\pi, 0)$. In light of the above discussion, $\bk_1$ and $\bk_2$ ($\bk_3$ and $\bk_4$) therefore both contribute to the Chern number by $+1$ (respectively $-1$) if Eq.~\eqref{eqn:limit} is satisfied. As can easily be seen from Eq.~\eqref{eqn:behaviorAroundDiscontinuityPoints} above, we thus obtain $\nu_{\text{2D}} = +1 + 1 - 1 - 1 = 0$ for all values of $\beta$ except at the isolated points $\beta = 0$ and $\beta = \pm 4$ where the dissipative gap closes.

We finally indicate the symmetry class of the 2D bulk system according to the Hamiltonian classification of Ref.~\cite{Ryu10}. Such a classification is made possible here due to the fact that the steady state of the translation-invariant system is pure, and thus identical to the ground state of the parent Hamiltonian $H_\text{parent}$. Since the latter only has particle-hole symmetry (C) with unitary operator $U_C = \sigma_x$, the 2D model belongs to the D class.

\subsection{1D case: winding number $\nu_{\text{1D}}$}

We now examine the case of an infinite, translation-invariant 1D system with Lindblad operators as defined by Eq. (4) in the main text, namely,
\begin{align}
	L_i = \beta' \, a^\dagger_i + ( a^\dagger_{i-1} + a^\dagger_{i+1} ) + ( -a_{i-1} + a_{i+1} ),
\end{align}
or, in momentum space,
\begin{align}
	L_k = 2 \mathrm{i} \sin{(k)} a_k + (\beta' + 2 \cos{(k)}) a^\dagger_{-k}.
\end{align}
As in the 2D case discussed above, the dissipative gap is proportional to the minimum of the norm $\mathcal{N}_k \equiv \{ L^\dagger_k, L_k \} = \beta'^2 + 4 \beta' \cos{(k)} + 4$. Since the latter vanishes for $\beta' = \pm 2$ (at $k = \pi$ and $k = 0$, respectively), we must distinguish three regions in parameter space, namely: $\beta' < -2$, $| \beta' | < 2$, and $\beta' > 2$. As long as $| \beta' | \neq 2$, we can define normalized Lindblad operators $L_k \to L_k / \sqrt{\mathcal{N}_k} = u_k a_k + v_k a^\dagger_{-k}$, with
\begin{align}
	u_k & = \frac{2 \mathrm{i} \sin{(k)}}{\sqrt{\beta'^2 + 4 \beta' \cos{(k)} + 4}}, \\
	v_k & = \frac{\beta' + 2 \cos{(k)}}{\sqrt{\beta'^2 + 4 \beta' \cos{(k)} + 4}},
\end{align}
and represent the latter by a real unit vector $\vect{n}_k = (n^x_k, n^y_k, n^z_k) \equiv (0, - 2 u_k \operatorname{Im} v_k, u_k^2 - | v_k |^2)$, which can be cast into the form of a complex number $e^{i \phi_k} = n^z_k + \mathrm{i} n^y_k$ for some angle $\phi_k$. Based on this construction, the so-called \emph{winding number} topological invariant then takes the form~\cite{Ryu02, Diehl11}
\begin{align} \label{eqn:windingNumber}
	\nu_{\text{1D}} & \equiv \frac{1}{2\pi} \int_{\mathrm{BZ}} dk \; \vect{a} \cdot ( \vect{n}_k \times \partial_k \vect{n}_k ) \nonumber \\
	& = \frac{1}{2\pi} \int_{\mathrm{BZ}} dk \, ( n_k^z \frac{d n_k^y}{dk} - n_k^y \frac{d n_k^z}{dk} ) \nonumber \\
	& = \frac{1}{\pi} \int^\pi_0 dk \frac{d \phi_k}{dk} = \frac{1}{\pi} \int^\pi_0 d \phi_k \nonumber \\
	& = \frac{\phi_\pi - \phi_0}{\pi} \in \mathbb{Z}.
\end{align}
The associated topological phase diagram can be derived by examining the behavior around specific points located in the three parameter regions $\beta' < -2$, $| \beta' | < 2$ and $\beta' >2$ where the calculation is most simple and the topological features easiest to visualize. Here we find
\begin{align}
	\setlength{\extrarowheight}{3pt}
	\begin{array}{lll}
		\beta' \to -\infty: & u_k = -1; \; v_k = 0; \\
		& \vect{n}_k = (0, 0, 1); \; \phi_k = 0, \\
		\beta' = 0: & u_k = \cos{(k)}; \; v_k = -\mathrm{i} \sin{(k)}; \\
		& \vect{n}_k = (0, \sin{(2k)}, \cos{(2k)}); \; \phi_k = 2k, \\
		\beta' \to +\infty: & u_k = 1; \; v_k = 0; \\
		& \vect{n}_k = (0, 0, 1); \; \phi_k = 0.
	\end{array}
\end{align}
As given by Eq.~\eqref{eqn:windingNumber}, we thus obtain $\nu_{\text{1D}} = 2$ for $| \beta' | < 2$ and $\nu_{\text{1D}} = 0$ otherwise.

We finally remark that the 1D model is chiral and belongs to the BDI class. Using the fact that the bulk steady state is pure --as in the 2D case above-- this can easily be shown by examining the symmetries of the associated parent Hamiltonian. The latter has time-reversal symmetry (T) with unitary operator $U_T = 1$ (which trivially satisfies $U^*_T U_T = +1$) and particle-hole symmetry (C) with unitary $U_C = \sigma_x$ (which satisfies $U^*_C U_C = +1$), and thus indeed belongs to the BDI class.

\subsection{Mapping the 2D vortex model to a 1D wire}

We now provide further details regarding the mapping to a chiral 1D wire presented in the main text. For simplicity, we work in the continuum limit and set $\alpha = 1$, $\phi = 0$ in the general form of the cross Lindblad operators (Eq. (2) in the main text). In an annular region centered around the vortex core (with $r > r_c$; see main text), these operators take the form $L(x, y) = ((\beta + 4) + \partial^2_x + \partial^2_y) \, \psi^\dagger(x, y) + e^{-i \varphi} (\partial_x + \mathrm{i} \partial_y) \, \psi(x, y)$ where $\psi(x, y)$ denotes the fermionic field and $(x, y)$ are Cartesian coordinates on the plane. Expressed in polar coordinates $(r, \varphi)$ defined from the center of the vortex core, they become $L(r, \varphi) = ((\beta + 4) + \partial^2_r + \partial^2_\varphi / r^2 + (1 / r) \partial_r) \, \psi^\dagger(r, \varphi) + (\partial_r + \mathrm{i} \partial_\varphi  / r) \, \psi(r, \varphi)$, and the external $U(1)$ gauge field $e^{-i \varphi}$ describing the vortex phase does not appear explicitly anymore. The annular region can be mapped onto a cylinder with Cartesian coordinates $(x', y')$ by defining $x' = r, \, y' = r \varphi$ (see main text, Fig.~2(a), right). The Lindblad operators then transform as $L(x', y') = ((\beta + 4) + \partial^2_{x'} + \partial^2_{y'}) \, \psi^\dagger(x', y') + (\partial_{x'} + \mathrm{i} \partial_{y'}) \, \psi(x', y')$~\cite{Footnote1}, and we recover the form of Eq. (2) in the main text corresponding to our original translation-invariant model on the plane. This shows that there exists a one-to-one correspondence between our original model embedded in the plane with a single $\ell = 1$ vortex and the same model in cylinder geometry with no vortex (i.e. with $p$-wave operator $\partial_{x'} + \mathrm{i} \partial_{y'}$). Physically, this stems from the fact that the gauge field $e^{-i \varphi}$ that was imposed on the plane to describe the vortex naturally arises on the cylinder owing to the extrinsic curvature of the latter.

Starting from the cylinder model, one can perform a Fourier transform in the translation-invariant direction of the cylinder in order to decouple the 2D model into a stack of 1D wires. Two of them, associated with the modes corresponding to the momenta $0$ and $\pi$, turn out to exhibit parameter regimes in which they are topologically nontrivial, as indicated by the winding number topological invariant (see main text).

\section{Majorana zero modes in the chiral 1D wire}

We consider the extended 1D wire depicted in Fig. 1(c) in the main text, with position-dependent Lindblad operators of the form
\begin{align} \label{eqn:1DCrossLindbladOperators}
	L_i = \beta' \, a^\dagger_i & + a^\dagger_{i-1} + a^\dagger_{i+1} - f(r_{i-1}) a_{i-1} + f(r_{i+1}) a_{i+1},
\end{align}
where $f(r)$ denotes the vortex profile function (see main text) and $r$ the position along the wire (this notation emphasizes the fact that the 1D wire effectively describes the behavior of a dissipative vortex with vorticity $\ell = 1$ in the \emph{radial} direction away from its core, as discussed in the main text). Since we are mainly interested in the behavior of the original 2D system inside the vortex core, we assume that the wire is semi-infinite, with open boundary conditions on its left side imposed by the geometry of the vortex core alone, i.e. by the profile function $f(r)$. In the framework developed in the first section above, the Lindblad operators take the form of complex vectors $\vect{l}_i$ (defined such that $L_i = \vect{l}^T_i \vect{c}$ where $\vect{c}$ is a vector of local Majorana operators; see above) with non-zero components
\begin{align} \label{eqn:lindbladVectors}
	\begin{split}
		& (\vect{l}_i)_{2(i-1)-1} = (1 - f(r_{i-1}))/2 = \delta(r_{i-1}), \\
		& (\vect{l}_i)_{2(i-1)} = \mathrm{i}(-1 - f(r_{i-1}))/2 = -\mathrm{i}(1 - \delta(r_{i-1})), \\
		& (\vect{l}_i)_{2i-1} = \beta'/2, \\
		& (\vect{l}_i)_{2i} = -\mathrm{i}\beta'/2, \\
		& (\vect{l}_i)_{2(i+1)-1} = (1 + f(r_{i+1}))/2 = 1 - \delta(r_{i+1}), \\
		& (\vect{l}_i)_{2(i+1)} = \mathrm{i}(-1 + f(r_{i+1}))/2 = -\mathrm{i} \delta(r_{i+1}),
	\end{split}
\end{align}
where $\delta(r_i)$ is the deviation from the value $f(r_i) = 1$ found in the bulk, defined such that $f(r_i) = 1 - 2 \delta(r_i)$ with $0 \leq \delta(r_i) \leq \tfrac{1}{2}$. As shown previously, a real unit vector $\vect{v}$ defines a Majorana zero mode $\gamma = \vect{v}^T \vect{c}$ if and only if $\vect{l}^T_i \vect{v} = 0$ for all $i$. Noticing that the components $(\vect{l}_i)_j$ are here real (respectively purely imaginary) for odd (even) indices $j$, we can restrict our search for Majorana zero modes to vectors $\vect{v}$ of the form
\begin{align}
	\text{(i) } & (\vect{v})_{2j-1} = \alpha_j, (\vect{v})_{2j} = 0, \label{eqn:case1} \\
	\text{or (ii) } & (\vect{v})_{2j-1} = 0, (\vect{v})_{2j} = \alpha_{j}. \label{eqn:case2}
\end{align}
with $\alpha_j \in \mathbb{R}$. As will be argued later, both cases give similar results owing to the symmetry of Eq.~\eqref{eqn:lindbladVectors} when considering even and odd components separately. We thus focus on (i), in which case the necessary and sufficient condition $\vect{l}^T_i \vect{v} = 0$ translates as
\begin{align} \label{eqn:recurrenceRelation}
	\alpha_{i+1} (1 - \delta(r_{i+1})) + \alpha_i (\beta'/2) + \alpha_{i-1} \delta(r_{i-1}) = 0.
\end{align}
In general, such a homogeneous linear second-order recurrence relation cannot be solved in closed form since its coefficients are not constant. In order to make progress, we examine the possible solutions away from the vortex core (i.e. away from the left edge of the 1D wire) where the vortex profile function is constant, namely, $\delta(r_i) = \delta$. In that case, one easily finds the general solution
\begin{align} \label{eqn:generalSolution}
	\alpha_i = C_{+} (r_{+})^i + C_{-} (r_{-})^i,
\end{align}
with $C_{\pm} \in \mathbb{R}$ and
\begin{align}
	r_{\pm} = \frac{-\beta'/2 \pm \sqrt{(\beta'/2)^2 - 4 \delta (1 - \delta)}}{2 (1 - \delta)}.
\end{align}
Since $r_{\pm}$ must be real, this solution exists if and only if $\abs{\beta'} \geq 4 \sqrt{\delta (1 - \delta)}$, which is trivially satisfied away from the vortex core where $f(r_i) = 1$, i.e. $\delta = 0$. The general solution then reads $\alpha_i = C (-\beta'/2)^i$ with $C \in \mathbb{R}$, and is normalizable if and only if $\abs{\beta'} < 2$. Assuming that this condition is satisfied, the solution can be extended inside the vortex core using the recurrence relation of Eq.~\eqref{eqn:recurrenceRelation}, \emph{provided} that the extension is consistent with the boundary conditions imposed at the center of the vortex core (i.e. on the left edge of the wire). Owing to the rotational symmetry of the vortex, the boundary conditions must be defined as
\begin{align}
	\alpha_{-1} = \alpha_{1}, \quad \delta(r_{-1}) = \delta(r_{1}),
\end{align}
where we have chosen the index of the leftmost site of the wire as $i = 0$. Using Eq.~\eqref{eqn:recurrenceRelation}, we then find $\alpha_{1} = \alpha_{0} (-\beta'/2)$. We must therefore have $\alpha_{0} > 0$ in order to obtain a non-trivial solution. Since $\vect{v}$ must be normalized to unity, there is no freedom in choosing the value of $\alpha_{0}$, and the solution of Eq.~\eqref{eqn:recurrenceRelation} is \emph{unique}. This proves that there exists a \emph{single} Majorana zero mode localized in the vortex core (or, equivalently, on the edge of the 1D wire) in the whole parameter range $\abs{\beta'} < 2$, and that the latter decays exponentially away from the vortex core, on a characteristic length scale $\xi = \xi(\abs{\beta'}) \sim 1/\abs{\log{(\abs{\beta'}/2)}}$ which diverges as $\abs{\beta'} \to 2$, as argued in the main text (and as verified numerically).

Remembering that the Chern number vanishes for $\abs{\beta'} < 2$ (i.e. for $0 < \abs{\beta} < 4$; see main text), the existence of a single Majorana zero mode in the vortex core is somewhat surprising: according to the well-known bulk-edge correspondence for the Chern number~\cite{Hatsugai93, Kitaev06, Ryu10, Gurarie11}, one would naively expect the existence of an \emph{even} number of such modes. We wish to emphasize that there is no contradiction here, since bulk-edge correspondence arguments only apply to the parent Hamiltonian $H_\text{parent} = \sum_i L^\dagger_i L_i$ (or to the matrix $\mathrm{i} Y$) defined above; not to the damping matrix $X$ which solely determines the existence of Majorana zero modes in our dissipative setting (see Eq.~\eqref{eqn:XYMatrices} and discussion thereof). In the present case, the bulk spectrum and eigenmodes of $H_\text{parent}$ exactly coincide with those of $X$ due to the fact that the Lindblad operators anticommute with each other in the bulk. However, the fact that these anticommutation relations are not satisfied in the vortex core leads to a loss of purity. In light of the general discussion presented in the first section above, we expect the bulk-edge correspondence pertaining to $H_\text{parent} = \sum_i L^\dagger_i L_i$ to be reflected in the appearance of a purity zero mode, as seen numerically (see main text).

We finally remark on the case given by Eq.~\eqref{eqn:case2} which has not been considered so far. Noticing in Eq.~\eqref{eqn:lindbladVectors} that the odd components $(\vect{l}_i)_{2j-1}$ --for increasing $j$-- have the same form as the even components $(\vect{l}_i)_{2j}$ --for decreasing $j$-- one can easily convince oneself that Eq.~\eqref{eqn:case2} corresponds to the symmetric situation of a single Majorana zero mode localized on the \emph{right} side of the 1D wire. This can be shown explicitly by imposing open \emph{dissipative} boundary conditions on the right of the 1D wire (thereby introducing an edge in the original 2D system) and repeating the construction above. Since such dissipative boundary condition appearing on (and defining) the physical edge of the system far away from the vortex core are mathematically similar to those appearing in the vortex core, a single Majorana zero mode must also be found on the edge of the system. In general, these boundary conditions will not be compatible with the Majorana zero mode localized in the vortex core; similarly, the boundary conditions imposed in the vortex core will not be compatible with the Majorana zero mode found on the edge of the system. As a result, both Majorana zero modes will acquire a small damping rate $\lambda \sim (\abs{\beta'}/2)^{2N}$ which decreases exponentially with the system size $N$, i.e. with the distance between them. The situation is therefore reminiscent of that of interacting Majorana modes in a Hamiltonian system: $\abs{\beta'}$ can be viewed as a coupling constant and $\lambda$ as a small splitting of the Majorana modes caused by inter-edge tunneling.

\section{Implementation in ultracold atom systems}

Here we discuss how to implement the dissipative dynamics investigated in the main text (see e.g. Eq. (2)) with spinless fermionic atoms in optical lattices. In ultracold atom experiments, spin is not a fundamental quantity, but is realized by a specific hyperfine state. If only a single hyperfine level of the atoms is populated, one obtains an ensemble of fermions that are effectively spinless. This underlies the possibility of realizing the Kitaev's Majorana wire with cold atoms~\cite{Jiang11}. The idea of the implementation of the situation with no vortex follows Refs.~\cite{Diehl08, Diehl11}; we summarize its main ingredients below. We first sketch how microscopically number-conserving Lindblad operators, corresponding to a quartic Liouville operator, relate to the quadratic master equation of the main text in the long-time limit. We then show how to implement the number-conserving operators in practice, and in particular how to optically imprint a dissipative vortex as defined in the main text.

\subsection{From number-conserving quartic to quadratic dissipative dynamics}

We start from a dissipative dynamics that microscopically \emph{conserves particle number},
\begin{align} \label{eqn:micro}
	\partial_t \rho = \kappa_0 \sum_{i=1}^N \left( \ell_i \rho \ell_i^\dagger - \tfrac{1}{2} \{ \ell^\dagger_i  \ell_i, \rho \} \right),
\end{align}
with bilinear Lindblad operators of the form $\ell_i = C^\dagger_i A_i$, where $A_i$ ($C_i^\dagger$) is a quasi-local superposition of fermionic annihilation (creation) operators. As anticipated in the main text, linear Lindblad operators such as the ones of Eq. (2) (see main text) then arise in the long-time and thermodynamic limit. More precisely, within a dissipative analog of BCS mean-field theory controlled by the proximity to the exactly known steady state (see the appendix of Ref.~\cite{Diehl11} for a detailed discussion), one can show that
\begin{align} \label{eqn:ellLrel}
	\ell_i = C^\dagger_i A_i \stackrel{t \to \infty}{\longrightarrow} L_i =  C^\dagger_i + \alpha e^{\mathrm{i} \phi} A_i
\end{align}
for spinless fermions. This results in the following mean-field master equation governing the late-time damping of excitations close to the steady state:
\begin{align} \label{eqn:meanfield}
	\partial_t \rho = \kappa \sum_{i=1}^N \left( L_i \rho L_i^\dagger - \tfrac{1}{2} \{ L^\dagger_i  L_i, \rho \} \right).
\end{align}
In Eq.~\eqref{eqn:ellLrel}, the phase $\phi$ originates from spontaneous $U(1)$ symmetry breaking, i.e., it is not fixed by the dissipative dynamics~\cite{Yi12}; it will be set to zero in the following, without loss of generality. The relative strength $\alpha > 0$ between creation and annihilation parts, on the other hand, is fixed by the system filling according to a specific number equation (see Eq.~\eqref{eqn:number} below). For the homogeneous setting in the thermodynamic limit, this number equation --as well as further characteristics of the late-time dissipative dynamics-- can be specified explicitly. Using similar notations as before (see e.g. Eq.~\eqref{eqn:Lops}), the Fourier transform of Eq.~\eqref{eqn:meanfield} can be written as
\begin{align} \label{eqn:meanfieldq}
	\partial_t \rho = \int \frac{d^2 \bk}{(2\pi)^2} \kappa_\bk \left( \bar{L}_\bk \rho \bar{L}_\bk^\dagger - \tfrac{1}{2} \{ \bar{L}^\dagger_\bk  \bar{L}_\bk, \rho\} \right),
\end{align}
with~\cite{Footnote2}
\begin{align} \label{eqn:qdefs}
	\bar{L}_\bk & = \mathcal{N}^{-1/2}_\bk L_\bk, \quad L_\bk = ( u_\bk a_\bk + \alpha v_\bk a^\dagger_{-\bk} ), \nonumber \\
	\mathcal{N}_\bk & = |u_\bk|^2 +  |\alpha v_\bk|^2, \quad \kappa_\bk = \kappa \, \mathcal{N}_\bk.
\end{align}
The explicit form of $u_\bk$, $v_\bk$ relevant to our model is given by Eq.~\eqref{eqn:Lops}. Note that we have kept the coefficient $\alpha$ explicitly (as opposed to setting $\alpha = 1$ as in Eq.~\eqref{eqn:crossLindbladOperators} for example) in order to examine how it relates to density. The fact that the pair wavefunction $\varphi_\bk = v_\bk / u_\bk$ is antisymmetric under momentum reflection (i.e. $\varphi_\bk = - \varphi_{-\bk}$) implies that $\{ \bar{L}_\bk , \bar{L}_\bq \} = \{ \bar{L}^\dagger_\bk , \bar{L}^\dagger_\bq \} = 0$ for any pair of momenta $\bk$, $\bq$. In addition, the normalization chosen in Eq.~\eqref{eqn:qdefs} guarantees that $\{ \bar{L}_\bk , \bar{L}^\dagger_\bq \} = \delta(\bq - \bk)$. The normalized Lindblad operators are thus proper fermionic quasiparticle operators; they generate a full Dirac algebra for the system and determine the steady state uniquely.

In the homogeneous setting, the mean-field calculation yields a simple form for the number equation,
\begin{align} \label{eqn:number}
	n = \int_\text{BZ} \frac{d^2 \bq}{(2\pi)^2} \frac{|\alpha v_\bq|^2}{|u_\bq|^2 + |\alpha v_\bq|^2},
\end{align}
which explicitly relates the parameter $\alpha$ to a given fixed average particle number. The effective damping rate entering Eq.~\eqref{eqn:meanfield} can also be calculated explicitly; it relates to the microscopic one of Eq.~\eqref{eqn:micro} as
\begin{align} \label{eqn:dampingrel}
	\kappa  = \kappa_0 \int_\text{BZ} \frac{d^2 \bq}{(2\pi)^2} \frac{|u_\bq v_\bq|^2}{|u_\bq|^2+ |\alpha v_\bq|^2}.
\end{align}
The simplifications arising in the long-time limit of Eq.~\eqref{eqn:micro} are similar to the ones emerging in the low-energy limit of a Hamiltonian theory describing a superfluid. Indeed, the low-energy physics of a fermion theory with weak attractive interactions (i.e. a quartic fermion theory) is universally described by a quadratic BCS Hamiltonian. In our dissipative setting, the analog of the gap for single-particle excitations present in such theories is a \emph{dissipative gap} $\bar{\kappa} > 0$, defined as the minimum of the damping function $\kappa_\bk$, i.e. $\bar{\kappa} = \operatorname{min}_\bk \{ \kappa_\bk \}$. In the specific model considered in the main text (see Eq. (2)), the damping function features a gap for all values of $\beta$ except for $\beta = 0$, $\pm 4$. The dissipative gap closes at $\bk_* = (0, \pi), (\pi, 0)$ for $\beta = 0$, $\bk_* = (\pi, \pi)$ for $\beta = 4$, and $\bk_* = (0,0)$ for $\beta = -4$; in the vicinity of these points, the damping rate vanishes quadratically, i.e. $\kappa_\bk \sim (\bk - \bk_*)^2$. Physically, the existence of a dissipative gap (i) implies the exponentially fast convergence of all many-body observables to the steady state, and (ii) gives rise to a separation of time scales between the rapidly damped modes of the bulk and the (Majorana) zero-damping modes which are present for appropriate boundary conditions.

\subsection{Implementation of the number-conserving bilinear Lindblad operators}

The operators $\ell_i = C_i^\dagger A_i$ appearing in Eq.~\eqref{eqn:micro} can be engineered via two-step processes involving an auxiliary degree of freedom, as shown in Refs.~\cite{Diehl08, Diehl10, Diehl11}. Auxiliary degrees of freedom can be provided by the intermediate sites of an optical superlattice, giving rise to a second, excited Bloch band (the target system residing in the lower sites of the lattice corresponds to the lowest Bloch band). The first step consists in \emph{coherently} coupling a quasi-local superposition of fermions to the auxiliary degree of freedom, and is associated to the annihilation part $A_i$. The second step, on the other hand, introduces \emph{dissipation} in the form of spontaneous phonon emission, and relates to the creation part $C_i^\dagger$: a decay channel for the particles from the excited Bloch band back into the lowest one is opened by coupling to a reservoir into which the whole lattice system is immersed. A suitable reservoir is provided by a bosonic superfluid; the bosonic nature of the environment is particularly useful, as we will see below. Microscopically, the system-reservoir coupling can be obtained from standard $s$-wave density-density interactions $\sim \hat n_\text{sys}(x)\hat n_\text{bath}(x)$ between system and bath particles. Note that the resulting scattering processes conserve the number of particles in the system. Nevertheless, the release of system energy necessary to decay spontaneously from the excited to the lowest Bloch band is only possible via the emission of phonons into the superfluid reservoir: energy exchange with the reservoir therefore occurs, while particle exchange does not. Ultimately, the coherent coupling to the auxiliary sites can be operated in such a way (far detuned) that the latter can be integrated out. We then end up with an effective dissipative dynamics for the lowest Bloch band only, described by operators $\ell_i = C^\dagger_i A_i$ --the annihilation part $A_i$ results from the coherent coupling to the auxiliary site, and the creation part $C_i^\dagger$ from the spontaneous decay back to the target Bloch band. For a detailed exposition of this physics, we refer to~\cite{Muller12}.

We now examine how to implement the specific annihilation and creation parts relevant to our problem. $A_i$ is obtained from coherent driving, i.e. $A_i = \sum_j \Omega_{ij} a_j$, where $\Omega_{ij}$ is the Rabi frequency connecting to the auxiliary site. This quantity is thus fully determined by the properties of the coherent drive laser. In particular, the short-distance $p$-wave structure of $A_i$ can be realized with specific wavelength ratios and relative phases between the laser generating the optical lattice and the drive laser. The longer-range vortex profile, on the other hand, can be implemented with a drive laser carrying an optical vortex; such vortices are easily obtained e.g. in the far-field regime of a fork hologram or a spiral phase plate~\cite{Heckenberg92, Allen92, Molina07}. This results in an annihilation part with Rabi frequencies of the form
\begin{align}
	\Omega_{ij} \sim f(r_j) e^{\mathrm{i} \ell \varphi_j} \left( \delta_{i - \vect{e}_x, j} - \delta_{i + \vect{e}_x, j} + \mathrm{i} ( \delta_{i - \vect{e}_y, j} - \delta_{i + \vect{e}_y, j} ) \right), \nonumber
\end{align}
where $f(r)$ is a slowly varying profile function proportional to the laser amplitude encoding the optical vortex, e.g. for a Laguerre-Gauss beam $f(r) \sim (r/l) e^{-(r/\xi)^2}$ with $l, \xi \gg a$ ($a$ being the lattice spacing). In this way, the optical angular momentum of light is directly imprinted onto the matter system. 

In contrast, the creation part results from spontaneous emission from the auxiliary site, and is therefore generically isotropic. This gives $C^\dagger_i$ the desired $s$-wave symmetry (see Eq. (2) in the main text). The spatial extent of $C_i^\dagger$, as well as the relative strength of the phase-transition parameter $\beta$, can be tuned through the bath properties~\cite{Diehl08}. With $A_i = \sum_j \Omega_{ij} a_j$ encoding the vortex phase and profile, the form of the Lindblad operators associated with a dissipative vortex therefore precisely matches the one introduced in the corresponding section on p. 2 in the main text; it reduces to the desired translation-invariant form of Eq. (2) (main text) for constant Rabi frequency. We have verified numerically that the precise structure of the vortex (different choices of $f(r)$) does not affect the findings presented in the main text.

Finally, it is clear that configurations of several vortices can be produced optically, and that single vortices can in principle be moved adiabatically around the lattice by displacing the optical device generating the optical vortex. Since (i) adiabatic parameter changes of the Liouville operator give rise to a gauge transformation just as in a Hamiltonian dynamics~\cite{Diehl11} and (ii) the vortices typically carry single (unpaired) Majorana modes (see main text), the exchange statistics must be non-Abelian~\cite{Ivanov01}.

\subsection{Discussion}

We comment on the conservation of particle number and parity in our system. A first key point of our setting is that it relies microscopically on the dissipative coupling to a \emph{bosonic} environment --in contrast to many solid-state implementations of fermionic superfluids or superconductors-- such that the fermion parity of the target system is an excellent quantum number: any coupling of bosonic degrees of freedom to fermions must be \emph{even} in the number of fermion operators due to superselection rules. This is an important prerequisite for robust topological order~\cite{Kitaev00, Akhmerov10}. Second, the coupling to the reservoir is such that the fermionic particle number is exactly conserved microscopically (that is, $[\ell_i, \hat{N}] = 0$ for all $i$, where $\hat{N} = \sum_i \hat{n}_i$ is the total fermionic particle number), which obviously implies exact parity conservation. Exact number conservation is given up subsequently in the mean-field theory, in full analogy to the BCS strategy, and justified by the equivalence of fixed-number and fixed-phase BCS states in the thermodynamic limit, where relative number fluctuations in the fixed-phase BCS wavefunction scale as $\sim 1/\sqrt{N}$ ($N$ being the number of degrees of freedom). While it may seem as if particles are lost or created in the system due to the interaction with the environment from the linear Lindblad operators, this must really be understood as processes that annihilate or create fermions \emph{into or out of the fermionic pair mean field}. This is particularly transparent from the explicit form of the effective mean-field damping rate $\kappa$ of Eq.~\eqref{eqn:dampingrel}, which includes weighted integrals of correlation functions such as $\langle a_\bq a_{-\bq} \rangle \sim u_\bq v_\bq$ providing for a mean field for the mode $\bk$ (see Refs.~\cite{Diehl11, Yi12} for a detailed discussion). In particular, this entails that the parity of the fermionic target system is fixed via its initial state. Finally, we critically examine the boundary conditions in this light. The vortex is generated optically, and crucially does not modify the property of particle number conservation of the Lindblad operators. Therefore, no problem arises in vortices with particle or parity conservation. Particle number conservation at the edge of the system, in contrast, is a subtler issue. It could be achieved along the lines of Ref.~\cite{Jiang11}, but the need for such subtleties can be avoided by trapping any Majorana mode localized on the edge in an additional vortex.

